\documentclass{cernyrep}
\usepackage[T1]{fontenc}
\usepackage[bookmarks, colorlinks=true, linktoc=page, linkcolor=black, citecolor=black, urlcolor=blue]{hyperref}

\usepackage{fancyhdr}
\fancyhfoffset{4 mm}
\fancypagestyle{ARTTITLE}{%
\fancyhf{} 
\chead{\small Proceedings of the 2017 CERN--Accelerator--School course on \it Vacuum for Particle Accelerators, Glumsl\"ov, (Sweden)} 
\lfoot{Available online at \url{https://cas.web.cern.ch/previous-schools}}
\rfoot{\thepage\hspace*{3mm}}
 
}

\usepackage{varwidth}
\usepackage{xcolor}

\begin{document}

\title{Interaction\index{eled} Between Beams and Vacuum System Walls}

\author {R. Cimino}

\institute{Laboratori Nazionali di Frascati,  Frascati, Italy}

\begin{abstract}
In modern high-intensity accelerators, the circulating beam interacts in many ways with the vacuum beam pipe, causing a  variety of  different phenomena. Most of them have been discussed at length in other contributions to  this CAS report. I concentrate here on the effects associated with the presence of electrons in the accelerator beam pipes.
Low-energy electrons in accelerators are known to interact with the circulating beam,  giving raise to the formation of a so-called e$^-$ cloud.  e$^-$ cloud effects can be detrimental to beam quality and stability, especially for positively charged beams. I will first describe the origin and the basic features causing e$^-$ cloud formation in accelerators, which depend not only on the beam properties but mainly on the vacuum vessel surface properties. Such material properties and the way one can study them will be here briefly presented. Finally, some of the   e$^-$ cloud mitigation strategies adopted and proposed so far will be described and discussed.

\end{abstract}


\maketitle 
\thispagestyle{ARTTITLE}

%
%
%
%

\section{Introduction}
A circulating beam  interacts in many ways with the vacuum beam pipe.  For instance, an accelerated beam produces synchrotron radiation, which can directly deliver an additional heat load to the accelerator walls and may produce photoelectrons and induce gas desorption (either photon- or electron-stimulated). Also, the beam can generate ions by interaction with the residual gas still present in the vacuum vessel, and such ion flux can interact with the walls. All these phenomena may have detrimental effects, not only on accelerator vacuum performances but also on beam quality, being responsible for a number of beam instabilities. Some of those topics have been discussed at length in this school, and I  mainly concentrate my presentation and, therefore, my report here, on the effects associated with the presence of electrons in the accelerator beam pipes. All these phenomena are known as  e$^-$ cloud effects, and are here called ECEs  \cite{Cimino2014}.
The electron cloud phenomenon is a ubiquitous effect in  accelerators of positive particles. First observations of related effects date back to more than 50 years ago \cite{Budker66, Budker67, Zimmermann2004,Grobner1976}; since then many different phenomena observed in many accelerators have been related to the formation of an electron cloud.

\begin{figure}
\centering\includegraphics[width=.9\linewidth]{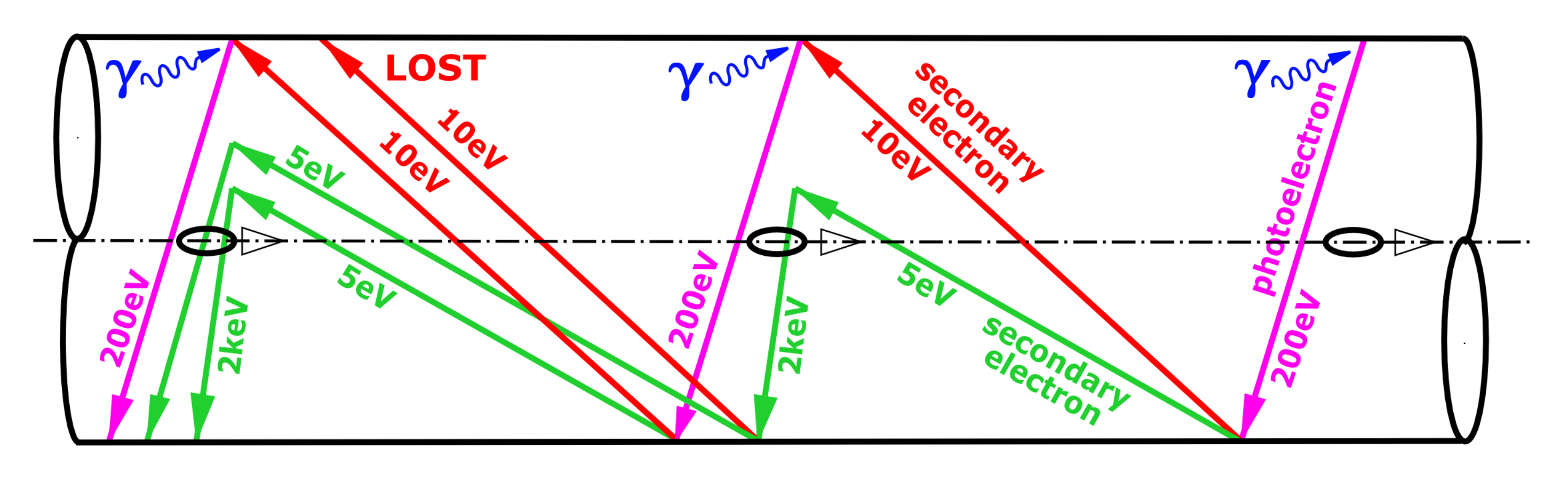}
\caption{e$^-$ cloud process (adapted from original representation by  F. Ruggero)}
\label{fig:figRuggiero}
\end{figure}

The phenomenology of the e$^-$cloud formation, together with an introduction to the most relevant  basic concepts, quantities, and tools, is analysed in Fig. \ref{fig:figRuggiero}, in which we slightly modify the artistic view of the e$^-$cloud phenomenon for the case of  the Large Hadron Collider (LHC) proposed by Francesco Ruggiero, head of the e$^-$ cloud crash program launched at CERN in 1997.

\begin{figure}
\centering\includegraphics[width=.9\linewidth]{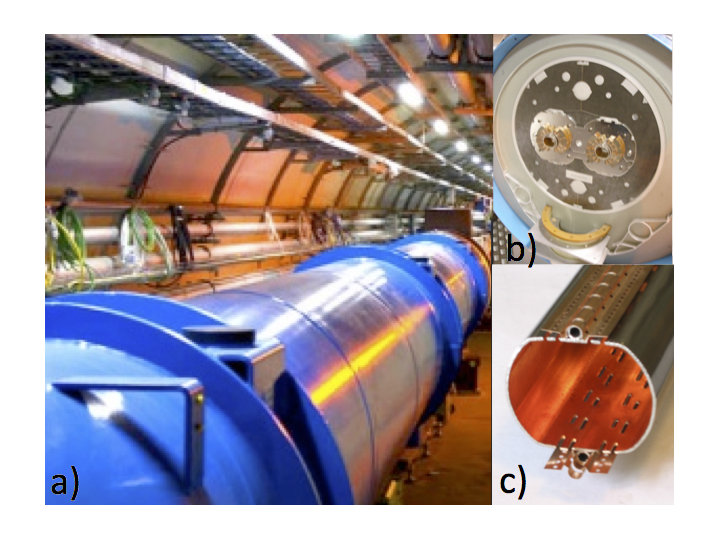}
\caption{(a) LHC tunnel; (b) inside of an LHC dipole; (c)  beam screen inside the cold bore of an LHC dipole \cite{Myers2013}}
\label{fig:LHCDipole}
\end{figure}

The LHC arc  dipoles have been described elsewhere \cite{Myers2013,LHC-DR} and are shown in Fig. \ref{fig:LHCDipole}. The interior of the dipole, see Fig. \ref{fig:LHCDipole}(b), is  a complex and highly technological realization. Briefly, it consists of a cold bore held at 1.9\,K  containing a so-called `beam screen', detailed in Fig. \ref{fig:LHCDipole}(c), held at a temperature between 5 and 20\,K, whose scope is essentially to protect the cold bore from unwanted heat deposition caused by image currents, synchrotron radiation,  \etc  The relevant LHC beam and chamber parameters can be found in the literature\cite{Cimino2014,Myers2013,LHC-DR}.

As soon as the first proton bunch appears in the dipole, see Fig. \ref{fig:figRuggiero},  being curved by the magnetic dipole field, it will emit synchrotron radiation.
These photons will travel tangentially to the beam orbit and hit the accelerator walls (around the orbital plane) with a grazing angle of incidence, about $1.5^\circ$ for the LHC case. At this point, they will be either reflected or absorbed by the wall. Some of the absorbed photons will produce photoelectrons, according to the wall surface photoelectron yield,  \ie the number of electrons produced per incident photon. The photoelectron yield, of course, will depend on many parameters, such as the actual photon energy, the angle of incidence,
or the surface roughness. Typically, photoelectrons have very low energy and, once emitted, travel very slowly in the accelerator beam pipe. Such photoelectrons will then interact with the positive proton beam.  It has to be noticed that photoelectrons are not the only possible seed to initiate the e$^-$ cloud process. Electron cloud effects have also been observed to occur in the Proton Synchrotron and  Super Proton Synchrotron
at CERN\cite{Arduini2001,rumolo2012}, where the synchrotron radiation produced by the  protons accelerated up to 25\,GeV and up to 450\,GeV, respectively, have way too little energy to generate photoelectrons.  Any electron, coming from residual gas ionization, is actually enough to initiate the process.
Once electrons are emitted from the accelerator walls, they will see the beam and, therefore, be accelerated by the Coulomb attraction. Some of them will be trapped by the beam but their limited velocity and their direction (being mainly normal to the beam and the walls) will then let them reach
the accelerator wall again, once the beam is passed,  as shown in  Fig. \ref{fig:figRuggiero}. During such interaction with the beam, they will gain energy, so that, once impinging against the opposite wall, they will generate a number of secondary electrons, whose intensity and energy distribution will depend on the beam pipe geometry, beam structure, and secondary electron yield (SEY)  of the accelerator's material. The SEY measures  the number of electrons produced per incident electron at a given energy; obviously, the higher it is, the higher will be the number of electrons multiplied at each bunch passage. This is a resonant phenomenon, since the next bunches will not only create new photoelectrons but also  accelerate the electrons already present in the vacuum system  towards the walls. At some stage, then, the e$^-$ cloud  will have such a high density as to influence the actual beam quality at each passage, inducing detrimental effects, up to its complete degradation and loss. Other detrimental effects may also accompany the build-up of the e$^-$ cloud, such as vacuum pressure rise or extra heat loads to the walls. The electron multiplication, called multipactoring, can be carefully studied by simulations,  which are an essential tool to address the e$^-$ cloud problem. A number of sophisticated computer simulation codes have been developed to study ECEs, and their predictions have been compared with experimental observations \cite{ECLOUDProc1, ECLOUDProc2, ECLOUDProc3, ECLOUDProc4, ECLOUDProc5}.
The quality and the prediction capability of such codes could be reached not only thanks to the efforts and capability of the scientists involved in their development, but also to their continuous benchmarking against different observations in working machines.  It is beyond the scope of this paper to discuss here such codes and their features; the interested reader is invited to look at the vast literature  existing on this topic\cite{ECLOUDProc1, ECLOUDProc2, ECLOUDProc3, ECLOUDProc4, ECLOUDProc5, Cimino2014,Cimino2015a,rumolo2012}.

During the process towards the understanding of the e$^-$ cloud phenomenon, it has been increasingly evident that an essential role is played by some relevant surface properties of the accelerator walls, calling for a huge experimental  effort to study such essential issues in dedicated laboratory experiments. Such studies  also gave some hints on how one could try to mitigate and control ECEs by dedicated  active or passive  mitigation strategies.

\section{Electron cloud effects on vacuum}

Electron cloud effects, as said, can be easily  accompanied by a pressure increase. Those vacuum effects will depend on the number of electrons hitting the wall, their kinetic energy, the wall surface material, \etc Moreover, since the e$^-$ cloud build-up is a non-linear mechanism, the pressure increase will usually show a clear threshold while varying beam intensity or structure \cite{Seeman2000,Bregliozzi2011}.
Since this contribution is within the CAS school on `Vacuum for particle accelerators', it is worth
showing one detailed example on how ECEs can affect vacuum. In the next section, an observed case for LHC is reported.

\subsection{An example: the LHC merged vacuum}

\begin{figure}
\centering
\includegraphics[width=.95\linewidth]{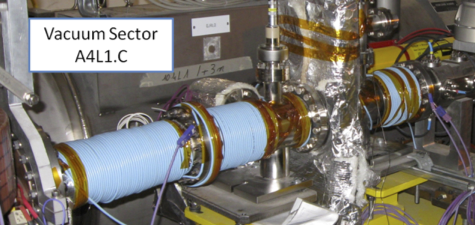}
\caption{Solenoids wrapped around the LHC merged vacuum beam pipe \cite{Bregliozzi2011}. (Courtesy of V. Baglin)}
\label{fig:mergedfoto}
\end{figure}

Figure \ref{fig:mergedfoto} shows a picture taken at the Vacuum Sector A4L1.C, where both proton beams circulate together. This implies that, in certain locations of this vacuum chamber, the double passing beam has a different time structure from that in other parts of the ring (when only one beam is circulating), giving rise to different resonance conditions that could trigger an ECE. A solenoid was wrapped around the vacuum chamber to add a Gaussian field, which will confine low-energy electrons close to the vacuum walls, stopping them from participating in e$^-$ cloud formation.

\begin{figure}
\centering\includegraphics[width=.95\linewidth]{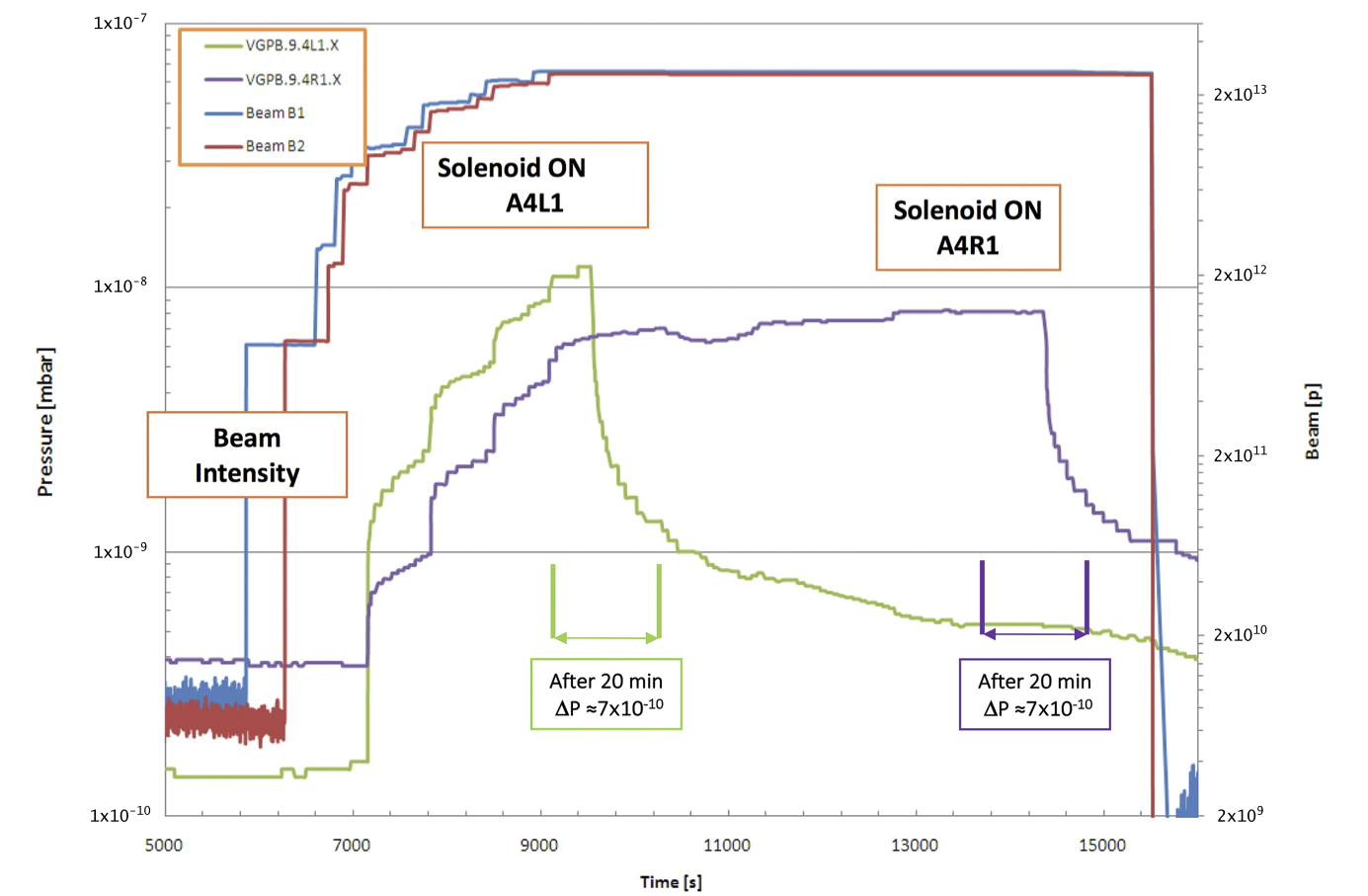}
\caption{Observed pressure reduction when a solenoidal field of 20\,G is applied. Blue and red lines represents countercirculating bunch population (right abscissa), green and purple are vacuum gauge readings (left abscissa) \cite{Bregliozzi2011}. (Courtesy of V. Baglin)}
\label{fig:Mergedvac}
\end{figure}

Figure \ref{fig:Mergedvac} shows the vacuum pressure readings (green and purple curves) during injection,  as measured in October 2010, during LHC commissioning phases. As soon as both proton beams were injected (see blue and red curves) a pressure run-away is clearly observed (above a certain threshold). When applying a solenoidal field of 20\,G, this pressure increase was quite rapidly reduced, down  to the $10^{-10}$\,mbar range. Indeed,  if the solenoidal field is switched off, ECEs and an associated pressure rise will appear again. A solenoid field could not have any potentiality in pumping a vacuum system; its only effect is to reduce the number of electrons participating in the cloud. Indeed, this is a very clear example to show how e$^-$ cloud formation can occur and affect a vacuum.

\section{Electron cloud build-up}

The electron cloud phenomenology described in the introduction shows how this phenomenon is a consequence of a number of effects and is strongly dependent on the beam and the actual accelerator material properties. Here, some essential steps necessary in e$^-$ cloud formation are briefly analysed.

\subsection{Primary electrons}

Primary electrons are generated during a bunch passage, both on the wall, as photoelectrons from synchrotron radiation (mainly for positron or electron beams) and as secondary electrons generated by beam loss (especially for ion beams) and, inside the beam volume, by the ionization of the residual gas. The number of electrons created per unit length by synchrotron radiation or by beam loss during one bunch passage can become comparable to the average line density of beam particles, in which case these processes alone can give rise to appreciable effects, called single bunch instabilities\cite{Ohmi2000}.
Without going into detail, such instabilities are important even if no resonant conditions triggers multipacting. The mere existence of a certain density of electrons in the vacuum chamber can affect beam stability, inducing unstable coherent motion of the whole bunch. This is especially true in machines with a high synchrotron radiation flux producing high photoelectrons densities; in those machines (\eg FCC-hh), extra care must be taken to analyse the formation of primary electrons as such even if not seeding any electron multiplication. Synchrotron radiation emission can be estimated starting from machine parameters (as described in many textbooks, \eg Ref.\cite{Jackson1975}) and the energetics of such an emitted spectrum are typically represented by the so-called `critical energy' $E_\mathrm{c}$, which is the photon energy that divides the total emitted energy into two regions of equal power. The minimum photon energy needed to create a photoelectron from an irradiated surface  depends on the material work function, and is typically between 4 and 5\,eV. For electron (positron) machines,  $E_\mathrm{c}$ can be as high as a few kiloelectronvolts, showing how such accelerators produce a high number of energetic photons, and, hence,  photoelectrons. Even in proton machines like LHC, it can be easily calculated that, at the nominal energy of $7$\,TeV, protons curved by a $8.4$\,T dipole field emit synchrotron radiation with a critical energy {$E_\mathrm{c}$} of about $44$\,eV \cite{baglin1998,cimino1999}. In the Future Circular Proton Collider (FCC-hh), a critical energy {$E_\mathrm{c}$} of about $4.5$\,keV is expected\cite{Dominguez2013}.

\subsection{Photons and photoelectrons}
\label{sec:primary}

\begin{figure}
\centering\includegraphics[width=.9\linewidth]{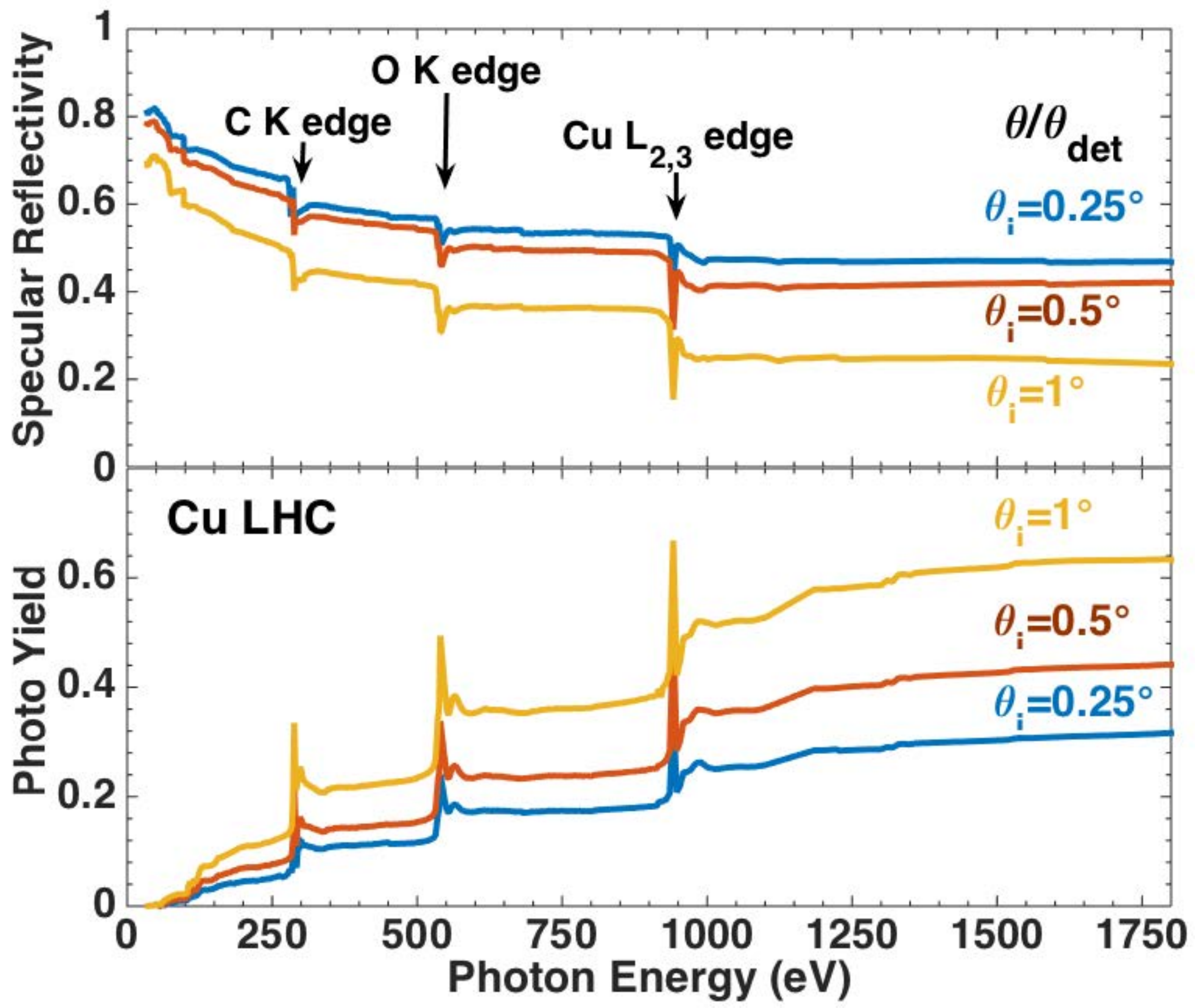}
\caption{Photoelectron yield (bottom panel) at different incidence angles and specular reflectivity (top panel) for various grazing incidence angles for an LHC Cu sample representative of the flat part of the beam screen, as a function of photon energy \cite{Eliana2017}.}
\label{fig:RPY}
\end{figure}

 In general, the photoelectron yield depends on the photon energy and angle of incidence and is a characteristic of the beam chamber material.
The azimuthal distribution of the photoemitted electrons depends not only on the beam parameters but also on the shape and reflective properties of the chamber wall.
The produced synchrotron radiation, for a given beam divergence, will illuminate the accelerator walls at a grazing incidence angle. Most of the photon beam will be scattered or reflected away; some will create photoelectrons, mostly in the presence of the dipole magnetic field perpendicular to the orbit plane but also outside the dipoles, in field-free regions. The electrons, photoemitted in the orbit plane, being affected by the magnetic field, are constrained to move along the field lines with a very small cyclotron radius;  thus, they will not be able to cross the vacuum chamber and participate in the e$^-$ cloud formation by gaining energy from the beam.  Conversely, the  reflected photon beam will soon illuminate top and bottom walls, emitting photoelectrons perpendicular to the orbit plane (hence, parallel to the magnetic field). Such photoelectrons will spiral along the field lines, efficiently participating in secondary electron production and, eventually, in multipacting as all the photoelectrons  are created in a field-free region. This simple reasoning shows how important it is to  experimentally determine the photon reflectivity of accelerator walls and its photoelectron yield \cite{cimino1999, Cimino2014, Dugan2015,Cimino2015a}. Recently  efforts have been made to use a state-of-the-art synchrotron radiation set-up to measure reflectivity and photoelectron yield as a function of monochromatic photon energy. An example of this kind of measurement is shown in Fig. \ref{fig:RPY}, where reflectivity and photon yield (\ie the number of electron produced during irradiation per incident photon) is measured on a LHC-like Cu surface as a function of photon energy and angle of incidence \cite{Cimino2014IPAC}.

Those data show that  the optical behaviour of a material is strongly dependent on its surface properties and surface quality. As observed in Fig. \ref{fig:RPY}, at the very grazing angle of incidence at which we are interested, the reflectivity and photoelectron yield of technical surfaces are not only influenced by the presence of adsorption edges typical of the Cu substrate (Cu $L_{2,3}$) but also, and quite significantly, by adsorption edge signatures of surface contaminants (O 1s and C 1s).

 Reflectometry can then be fruitfully performed
and the photon energy and the incidence angle dependence can give insight into the properties of the substrate and its contamination, while the amount of specular radiation
and the scattered light distribution give information on the optical quality of the surface.

\subsection{Energy gain}
\label{sec:egain}
Primary electrons  taking part in the e$^-$ cloud build-up are non-relativistic electrons ($v_\mathrm{e} \ll c$) accelerated in the field of a highly relativistic positive bunch ($\beta= v/c\approx 1$) moving in the longitudinal direction, $z$, of the beam pipe. The electric field of the relativistic bunch is Lorentz-contracted to a cone with an angle of order $1/\gamma$ in the direction perpendicular to the beam's motion. Thus, as a first approximation, the influence of the longitudinal kick produced by the beam can be neglected because its effect is small together with all the bunch-related magnetic effects.
In simple terms, one can distinguish two regimes as a function of the initial electron position\cite{Berg1997}: the electron is far from the bunch or it can get trapped in it. In both cases, the kick gained by the electron in the cloud after each beam passage can be calculated.

\begin{figure}
\centering\includegraphics[width=.95\linewidth]{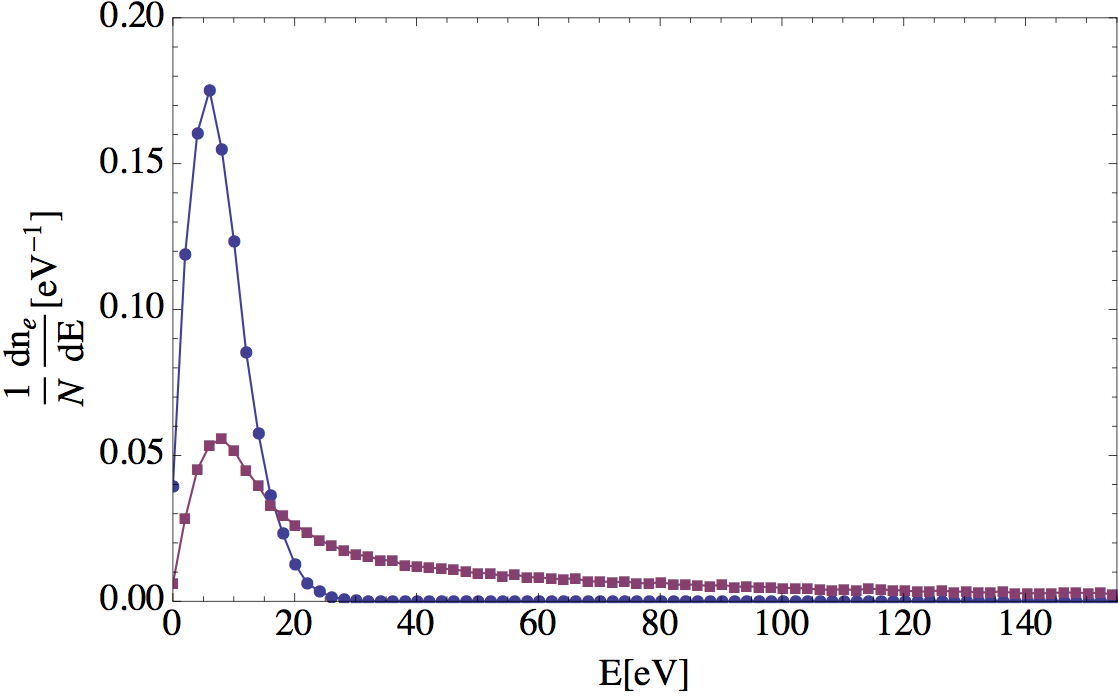}
\caption{Photoelectron energy distribution at the moment of emission (blue circles) and after the first bunch passage (red squares) as computed using the code ECLOUD \cite{ecloud-code} for an LHC dipole \cite{Cimino2014,Myers2013,LHC-DR}.}
\label{fig:EnGain}
\end{figure}

Figure \ref{fig:EnGain} shows an  estimate of  the electron energy distribution (red curve) gained after a single bunch passage, for an LHC dipole, as obtained using the code ECLOUD\cite{ecloud-code}. Calculations assume an  initial photoelectrons energy distribution  (blue curve in Fig. \ref{fig:EnGain}), which was modelled based on experimental findings \cite{cimino1999}, as a truncated Gaussian centred at 7\,eV, with a standard deviation of 5\,eV. The electrons in the chamber gain energy but maintain a significant density at very low energies. Than, once the proton beam is gone (see Fig. \ref{fig:figRuggiero}), such electrons drift towards the opposite wall of the accelerator chamber.

\subsection{Electron multipacting}
The primary mechanism causing a build-up of electrons is beam-induced multipacting. Here, electrons accelerated by the electric field of the passing bunches hit the vacuum-chamber wall with sufficient energy to produce, on average, more than one secondary electron per incident electron. The number of secondary electrons depends on the SEY of the chamber material, which is a function of the primary-electron energy, its angle of incidence, and the chamber surface composition and history.

\begin{figure}
\centering\includegraphics[width=.95\linewidth]{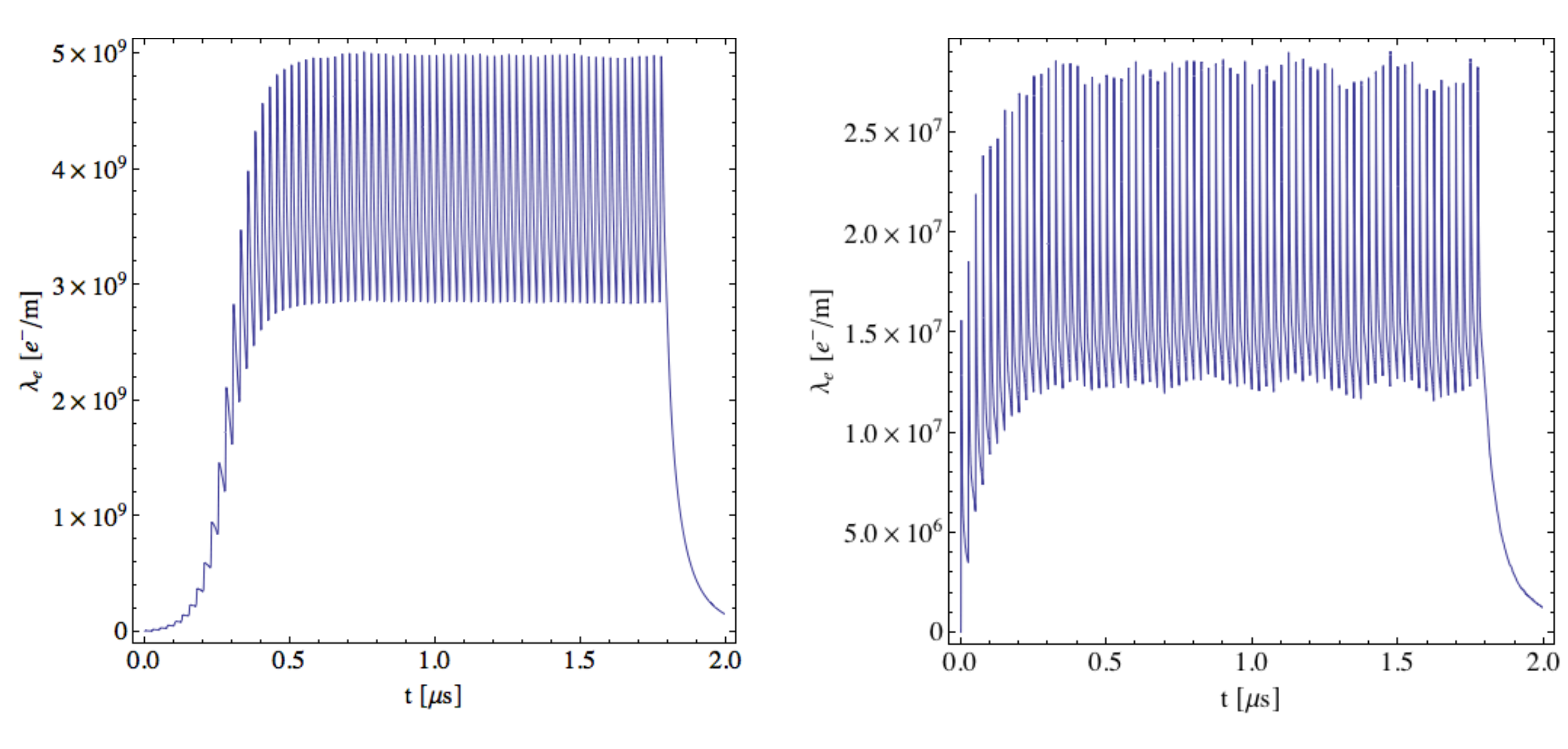}
\caption{Evolution of electron density computed using ECLOUD\cite{ecloud-code}. The case shown corresponds to a filling pattern featuring 72 charged bunches, with a bunch charge of $N=1.0\times10^{11}$ protons, followed by 28 empty (zero-charge) bunches. The assumed bunch spacing is 7.48\,m, and the maximum values of the SEY are $\delta_{\max}= 2.0$ (left) and $\delta_{\max}= 1.0$ (right).}
\label{fig:LHCbuildup}
\end{figure}

For a bunched beam, the induced multipacting could be interpreted as a resonance phenomenon. In particular, a multiplication of the secondary electrons takes place if, during the passage of a bunch, the electrons produced are accelerated and hit the chamber walls with an impact energy  able to produce more electrons than those originally created\cite{grobner, Cimino2014}.  In this way, after every bunch passage not only will new primary electrons be created and accelerated in order to produce secondary electrons, but also all the secondary electrons produced by previous bunches will be accelerated by the beam, leading to an exponential growth of the electron density.
However, the electron multiplication is not unlimited. At a certain level of the cloud density, the number of new electrons produced through the primary or secondary emission process will be equal to the number of electrons that are absorbed by the beam chamber, leading to a quasi-stationary equilibrium.  This levelling is clearly evident in Fig. \ref{fig:LHCbuildup}, where different electron densities are calculated in identical beam conditions, with  $\delta_{\max}$= 2.0  and 1.0, respectively. The  density at which this quasi-stationary equilibrium is reached depends strongly on the beam and chamber parameters and on the occurrence of the multipacting process. In general, if multipacting occurs, the e$^-$ cloud density grows exponentially until the equilibrium is reached under the influence of the space charge field of the cloud itself. Several analytical, as well as numerical, simulation models\cite{ECLOUDProc1, ECLOUDProc2, ECLOUDProc3, ECLOUDProc4, ECLOUDProc5} have been developed to obtain a deeper comprehension of the dependence of the e$^-$ cloud dynamics on the relevant parameters and to predict the ECE. Benchmarking  simulated results with machine observation allowed  simulation codes identifying crucial parameters to be refined. In this multiplication schema, SEY  becomes a surface parameter of paramount importance, and requires a more detailed analysis.

\subsection{Secondary electron yield}
The SEY measures the capability of a solid surface to produce secondary electrons, once it is  irradiated by electrons of different primary energy. It is commonly denoted $\delta(E)$ and has  been continuously  studied in  the accelerator community investigating issues related to  e$^-$ clouds\cite{ECLOUDProc1, ECLOUDProc2, ECLOUDProc3, ECLOUDProc4, ECLOUDProc5,Cimino2014}.
The SEY (\textit{$\delta(E)$}), is defined as the ratio  of the number of electrons leaving the sample surface (\textit{$I_\mathrm{out}(E)$}) to the number of incident electrons (\textit{$I_\mathrm{p}(E)$}) per unit area.
\textit{I$_\mathrm{out}(E)$}  is  the number of electrons emitted from the surface  but also the balance between the current flowing from the sample \textit{$I_\mathrm{s}(E)$} minus the current impinging on the sample, \textit{$I_\mathrm{p}(E)$}:
\begin{equation}
 \delta(E)=I_\mathrm{out}(E)/I_\mathrm{p}(E) =(I_\mathrm{p}(E) - I_\mathrm{s}(E) )/I_\mathrm{p}(E) = 1-I_\mathrm{s}(E)/I_\mathrm{p}(E).
\label{eq:delta_def}
\end{equation}

\begin{figure}
\centering\includegraphics[width=.95\linewidth]{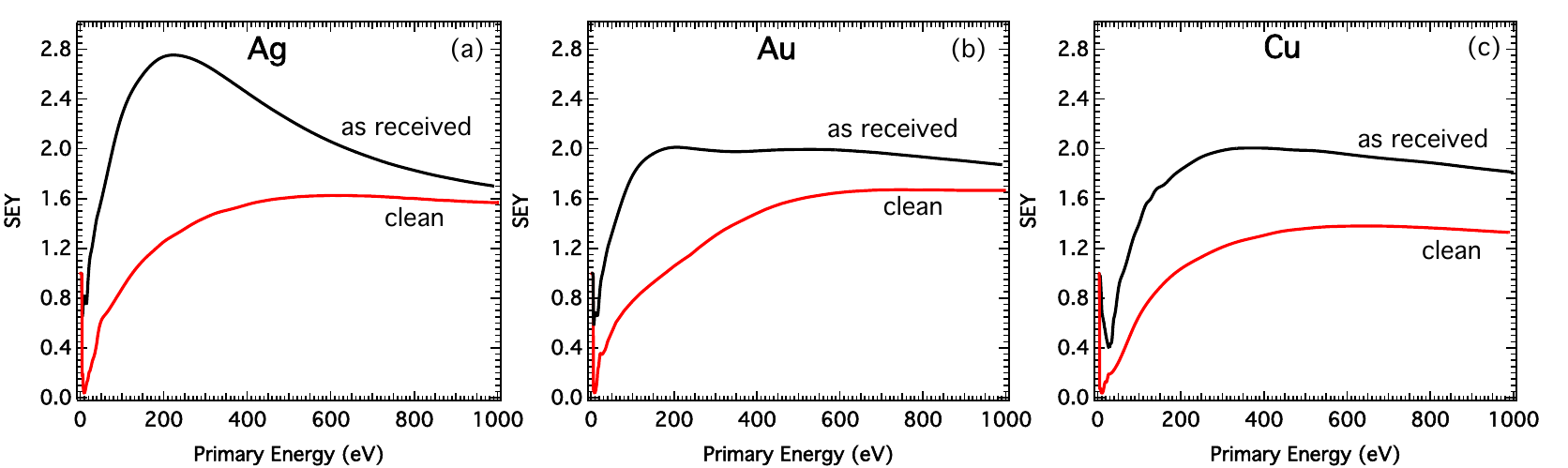}
\caption{Typical SEY curves measured for clean (black line) and as-received (red
line) surfaces of (a) Ag, (b) Au, (c) Cu polycrystalline samples. In all cases, the primary energy is
referred to the Fermi level\cite{Gomez2017}.}
\label{fig:SEYClean}
\end{figure}

To measure the SEY we have different experimental schemes: one can separately measure the impinging electron current generated by the electron source $I_\mathrm{p}(E)$ by means of a `Faraday cup', and then measure the sample current \textit{$I_\mathrm{s}(E)$}; one can simultaneously measure $I_\mathrm{s}(E)$  and $I_\mathrm{out}(E)$, collected either from a cage around the gun and placed in front of the sample or using a hemispherical collector all around the system. It is outside the scope of this contribution to analyse in detail all these different techniques. What matters here is to know that all methods, used with some care, produce very similar results.
The typical behaviour of the SEY as a function of the energy of the impinging electron is reported in Fig. \ref{fig:SEYClean}, where $E_{\max}$ is the value of the energy at which the SEY assumes its maximum value $\delta_{\max}$.   For the purpose of identifying significant parameters to compare different SEY curves, one can single out: (a) a low-energy region $(0\UeV\leq E_\mathrm{p} \leq 20\UeV),$ which is known as LE-SEY;
 (b) the energy value at which  $\delta(E)$  reaches its maximum:  $\delta_{\max}(E_{\max})$.
 In the following discussion, we will deal with the   $\delta_{\max}(E_{\max})$ of each surface, stressing the relevance of such parameter in the electron-cloud-related studies. However, it is worth remembering that it is the full dependence of the SEY  on the impinging primary energy,  not just its maximum value,
that does play an essential role in determining electron-cloud-related effects in accelerators.
 If most of the following discussion will deal with the  $\delta_{\max}(E_{\max})$ of each surface, we stress here that for electron-cloud-related studies  $\delta_{\max}(E_{\max})$ is indeed a relevant parameter, but it is the full dependence of SEY on the impinging primary energy that plays an essential role in  determining electron-cloud-related effects in accelerators.

\begin{figure}
\centering\includegraphics[width=.95\linewidth]{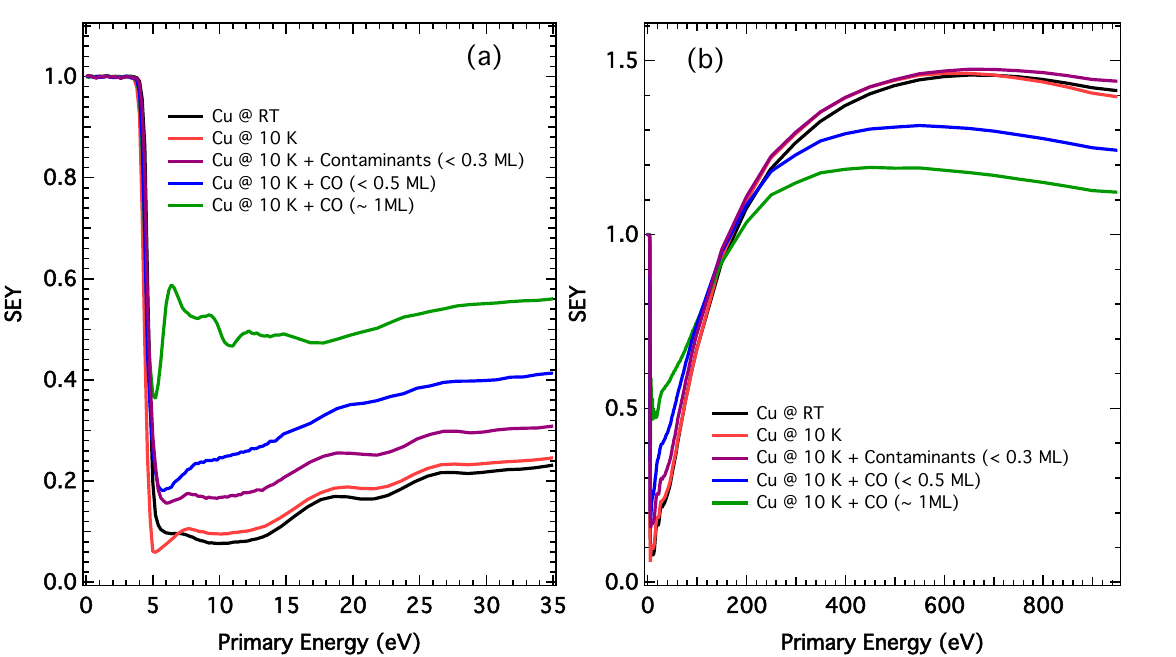}
\caption{(a) LE-SEY and (b) SEY curves measured at 10\,K on a polycrystalline Cu clean sample
(red) and in the presence of 0.3\,Ml of adsorbed residual gases (purple) and 0.5\,Ml (blue) and
1\,Ml (green) adsorbed CO. LE-SEY and SEY curves measured on the clean sample at room temperature
are shown for comparison (black) \cite{Gomez2017}. In all cases, the primary energy is referred to the Fermi level.}
\label{fig:SEYCont}
\end{figure}

The SEY depends not only on the wall material type but also, and somehow more significantly, by its surface composition. For the `as-received' technical materials normally studied in our accelerator vacuum context,  SEY can be only marginally related to the properties of the corresponding clean metal, being dominated by the presence of the surface contamination layer. This can be seen in Fig. \ref{fig:SEYClean}, where SEY curves for Ag, Au, and Cu polycrystalline samples were studied as both atomically clean and
`as-received' surfaces. It is clear that the measured SEY for a technical surface  depends on the contaminants more than on the underlying metal.  This is better illustrated in Fig. \ref{fig:SEYCont}, where LE-SEY and SEY curves are shown for an atomically clean Cu polycrystalline sample surface at 10\,K and for very small contamination levels.
Figure \ref{fig:SEYCont} provides strong evidence that the SEY (especially in its LE part) is so much sensitive to small
surface composition changes that the SEY itself could be used as a simple spectroscopic technique to trace the material surface condition\cite{Angelucci2017}.

\begin{figure}
\centering\includegraphics[width=.9\linewidth]{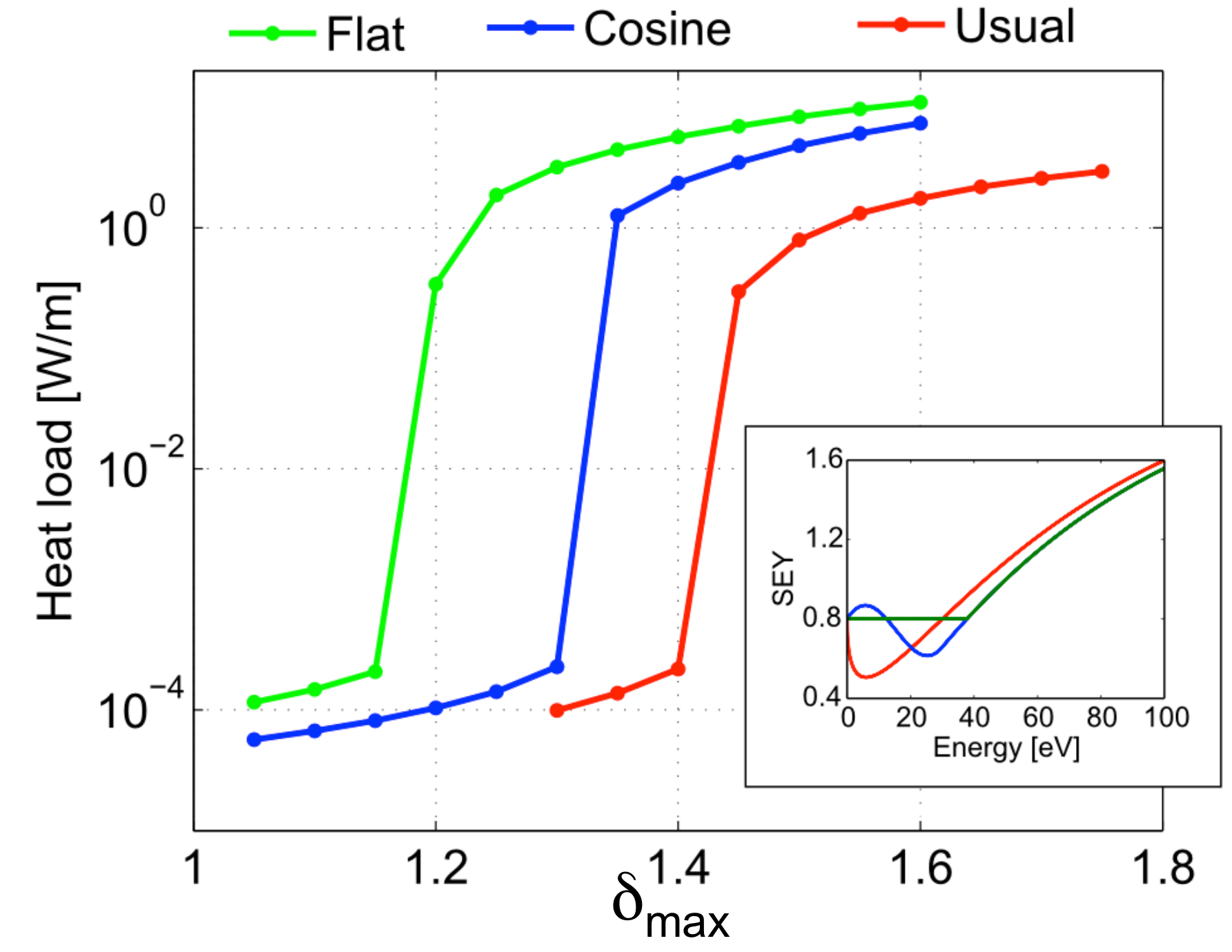}
\caption{Simulated \cite{ecloud-code} heat load as a function of $\delta_{\max}$ for the different LE-SEY behaviours shown in the inset. Inset: LE-SEY curve as parametrized in three different  ways: usual parametrization (red curve, higher $\delta_{\max}$ ); flat parametrization (green curve, lower  $\delta_{\max}$); cosine parametrization (blue curve) \cite{Cimino2015a}.}
\label{fig:HLxLESEY}
\end{figure}

The form and the value of the LE-SEY in the e$^-$ cloud build-up calculation have a direct impact on the results and should be studied with care. As said, calculations are performed by parametrizing the SEY curve as a function of
$\delta(E)$. Initially, its shape was defined to go to zero at zero primary electrons, since almost no LE-SEY was measured in the laboratory \cite{ECLOUDProc1, ECLOUDProc2, ECLOUDProc3, ECLOUDProc4, ECLOUDProc5} and the evidence supporting significant LE-SEY for technical surfaces \cite{Cimino04} was somewhat controversial \cite{Cimino2014,Cimino2015a}. Experimental evidence showing that a significant and certainly non-zero LE-SEY value must be expected from technical surfaces is now confirmed in many cases, as in  Figs. \ref{fig:SEYClean} and \ref{fig:SEYCont}. It is, therefore, important to run simulations that take into account a non-zero SEY at low energy. Figure \ref{fig:HLxLESEY} shows the heat load calculated  \cite{ecloud-code} by using the three different parametrization curves\cite{Cimino2015a}, as shown in the inset of Fig. \ref{fig:HLxLESEY}.

The different values  of the  ECEs onset and the corresponding different calculated heat load values, being related to ECE formation in LHC, by only changing the LE-SEY parametrization shows the importance of defining and measuring the LE-SEY and its stability versus operation.


Only by changing the LE-SEY parametrization are different values of $\delta(E)$ for the onset of the ECE  obtained, as well as the corresponding calculated heat load values, of great relevance to LHC operation. This evidence shows the importance of defining and measuring the LE-SEY and its stability versus operation.

In our context, it is expected that subtle surface modifications, as caused by electron- or photon-induced chemical surface changes  or the presence of physisorbed gases, as is usual on cold surfaces, could indeed have a direct effect on the SEY (and the LE-SEY) and hence on ECEs.
Also, as will be presented later, the SEY  depends on the surface morphology and a geometrically modified surface (at the nano-, micro-, or macroscale) will show a reduced SEY with respect to an equivalent flat surface. If those modification maintain the material compliant to all other   requirements (vacuum, impedance, \etc), surface roughening could be used to modify SEY at will.

\subsection{Cloud decay and electron survival}
\begin{figure}
\centering\includegraphics[width=.95\linewidth]{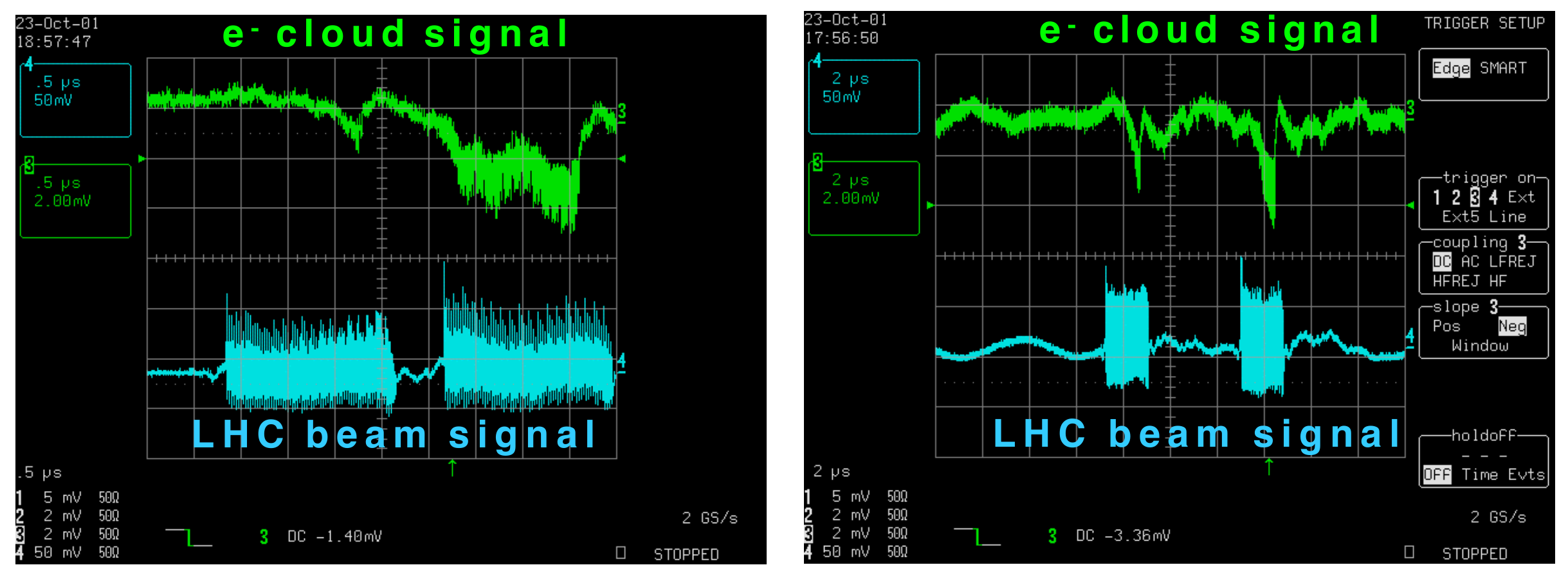}
\caption{Pick-up signal of the cloud build-up of two Super Proton Synchrotron batches composed of 72 bunches, spaced by 550 ns (left panel) and by $5.25\Uus$ (right panel) \cite{jimenez2002}.}
\label{fig:memory}
\end{figure}

Once the multiplication has started, as governed by SEY and other parameters, one way to moderate ECEs will be to put gaps between particle trains. On
so doing, the electron density will decay.  Experimental observations show that such decay is much slower than expected \cite{jimenez2002, Cimino2014} and a sort of memory effect seems to occur. This behaviour is shown in Fig. \ref{fig:memory}, where the pick-up signal of the cloud build-up of two Super Proton Synchrotron batches, composed of 72 bunches spaced by 550\,ns, shows that the cloud signal is much more ready to start during the passage of the second batch\cite{jimenez2002}. Several effects are suspected to contribute to the long memory and lifetime of the electron cloud: non-uniform fields, such as in quadrupoles or sextupoles, may act as magnetic bottles and trap electrons for an indefinite time period. It is now known that, on technical surfaces, low-energy electrons could have a SEY close to unity\cite{Cimino2015a}, so that they could survive for a long time, bouncing back and forth between the chamber walls, contributing to this observed effect.

\subsection{Simulations}
\begin{figure}
\centering\includegraphics[width=.95\linewidth]{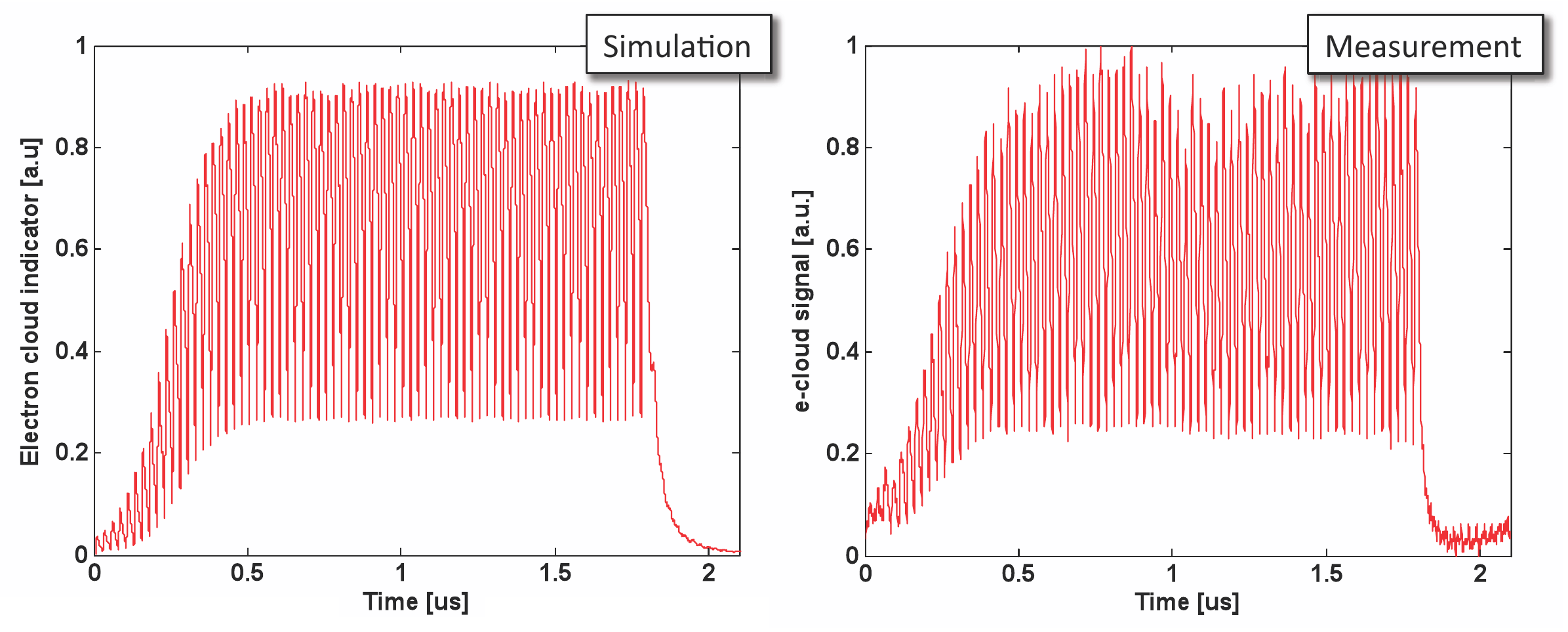}
\caption{e$^-$ cloud build-up simulation (left) and measurement (right) for a 25\,ns beam with $1.33 \times 10^{11}$ ppb, $4$\,ns long \cite{rumolo2012}.}
\label{fig:simul}
\end{figure}

It is not intended here to discuss simulation codes and activities that can be found in the literature and are only partially cited here\cite{ECLOUDProc1, ECLOUDProc2, ECLOUDProc3, ECLOUDProc4, ECLOUDProc5,Cimino2014,Cimino2015a,rumolo2012}. It is only important to state that the performed and ongoing simulation activity is fundamental to predicting ECEs on machine performances and that the reached level has been benchmarked against measurements with a very high confidence level. An example of this is shown in Fig. \ref{fig:simul}, which shows measured and simulated signals. The simulated signal is obtained with $\delta_{\max} = 1.6$.  The impressive resemblance between the two curves suggests that the assumed electron cloud model correctly describes the phenomenon, confirming how simulations performed with realistic parameters can match direct observations, as measured in real machines.

\section{Mitigation strategies}

Owing to the discussed importance of having a stable low SEY and, in some cases, a low photoelectron yield surface,  a number of mitigation strategies have been proposed  and studied. But there  is still    not a unique and final solution able to solve all problems related to a high SEY and, while R\&D is still actively ongoing, different mitigation strategies, or a combination of them, have been adopted and will be  briefly discussed here. Generally speaking, the countermeasures implemented today are of two kinds: passive strategies, which aim to reduce surface parameters, such as the SEY or photoelectron yield, and active strategies, which introduce external electric or magnetic fields to reduce the e$^-$ cloud formation. They will be presented briefly in the following,  trying to highlight the advantages and disadvantages of each method.

\subsection{Coatings}

\begin{figure}
\centering\includegraphics[width=.9\linewidth]{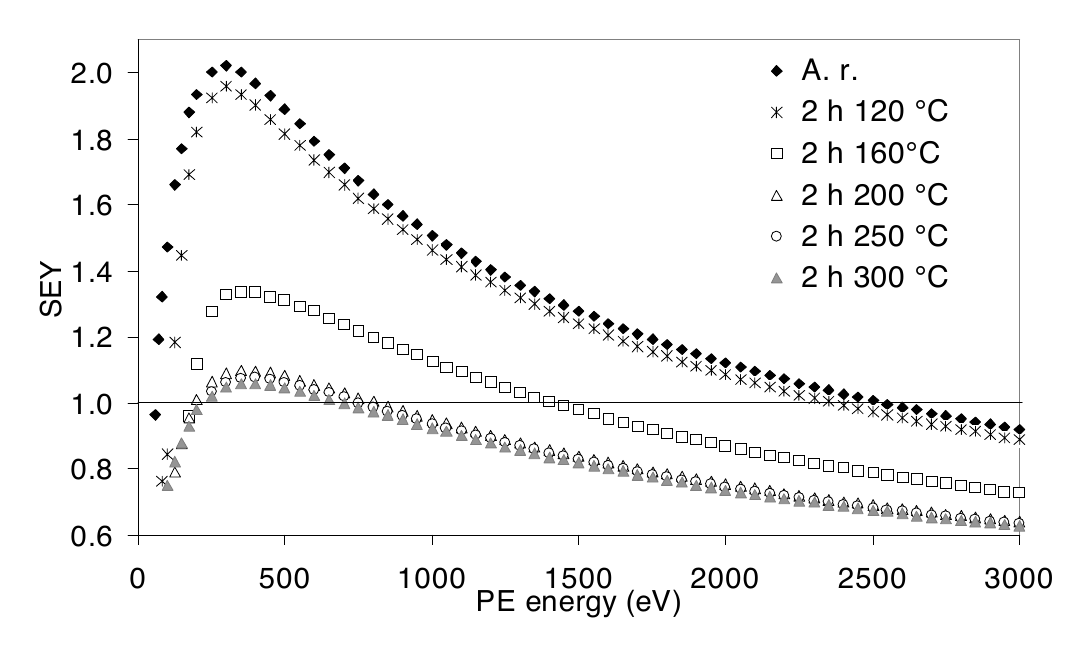}
\caption{SEY versus primary-electron (PE) energy of the TiZrV non-evaporable getter coating; as received (A.r.) and after 2\,h heating at 120 $^{\circ}$C, 160 $^{\circ}$C, 200 $^{\circ}$C, 250 $^{\circ}$C, 300 $^{\circ}$C\cite{Scheuerlein2000}.}
\label{fig:SEYNEG}
\end{figure}

Of course, the ideal solution to solve most of electron-cloud-related instabilities would be the identification of a material with intrinsically low and stable SEY, being compatible with all vacuum and impedance requirements. Such research has not yet given the desired results, in particular because, as discussed previously,  every material interacts with the atmosphere and is contaminated by it.  This unavoidable surface  layer renders most surfaces very similar to each other and coatings are unstable or easily contaminated.  This is indeed true for most metallic surfaces and coatings, with two significant exceptions, namely TiZV  non-evaporable getter (NEG) coating and amorphous
carbon. In the case of TiZrV NEG coatings, developed at CERN by Benvenuti  as diffuse pumps\cite{Benvenuti1999},  a mild thermal annealing in ultra-high vacuum (activation) is required, which has been demonstrated to remove  most of the surface contaminants \cite{Benvenuti1999, Scheuerlein2000, Chiggiato2006}, resulting in a stable and  close to atomically clean surface, with $\delta_{\max}$ about 1.1. This can be seen in Fig. \ref{fig:SEYNEG}, where the SEY is measured as a function of annealing at different temperature.  In this school,  Grabski from MAX IV discussed at length  the impedance budget issues connected with the diffuse use of NEG in the ring, so I refer to his work for further details.  In general, if ultra-high vacuum is required and if space allows to heat the accelerator wall at around 200 $^{\circ}$C to activate the NEG film, such coatings are a very efficient e$^-$cloud mitigator, as shown by the behaviour of all the NEG-coated warm sections of the LHC.
Of course, in magnets, it will be difficult to find space for heaters to activate the film and for the necessary insulators required to prevent damage
to the coils, so the benefit of SEY reduction and extra pumping given by the use of the NEG film should be balanced by considering the disadvantages related to an increase in the magnet bore, or alternatively, the reduction of  the  `in-vacuum' aperture \cite{Grabski2013}. Moreover, not only is the use of NEG films in cold environments more critical for designing and hosting the heaters
or insulators, but more studies should be performed on TiZrV NEG coating properties at low temperature since it is not yet known whether  such surface still pumps at low temperature and therefore whether it stays clean without physisorbed contaminants during operation.


\begin{figure}
\centering\includegraphics[width=.9\linewidth]{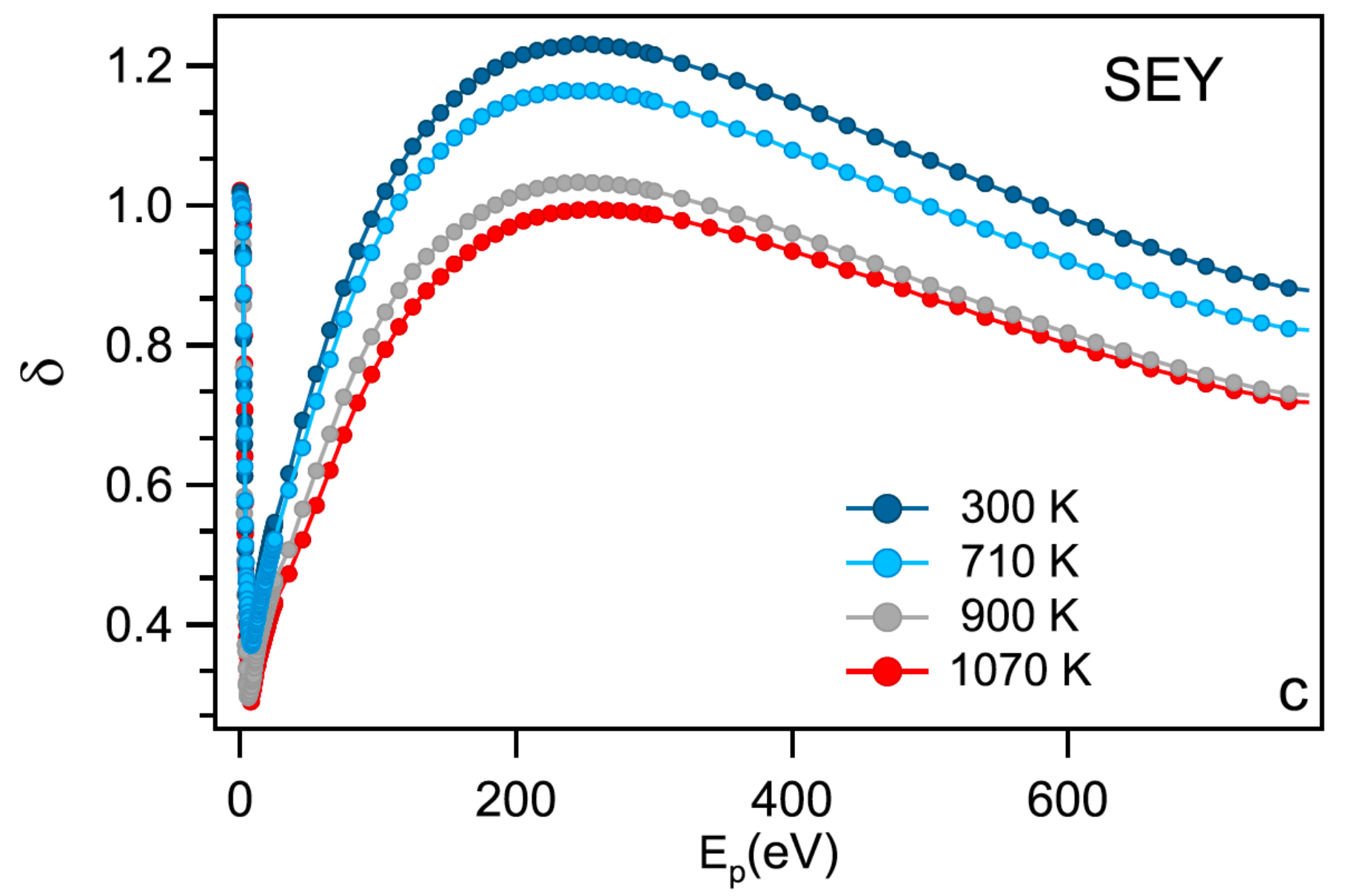}
\caption{SEY curves measured on  C films as a function of annealing temperature (Highest temperatures correspond to lower SEY)\cite{Larciprete2015}}
\label{fig:SEY-CvsT}
\end{figure}

Carbon, including its amorphous version, is also known to be quite unreactive, so it is expected not to adsorb significant contaminants that could modify its properties. It has been found  \cite{ECLOUDProc1, ECLOUDProc2, ECLOUDProc3, ECLOUDProc4, ECLOUDProc5,  Vallgren2011,Gonzalez2016,Larciprete2015} that graphite and amorphous carbon grown by magneto sputtering each form  an intrinsically low SEY material with $\delta_{\max}$  about 1.0. This low SEY depends on the detailed status of the sample and on its quality. A lower SEY is found in crystalline highly oriented pyrolytic graphite\cite{Gonzalez2016} and amorphous carbon \cite{Larciprete2015}. This is shown in Fig. \ref{fig:SEY-CvsT}, where SEY curves for amorphous carbon are shown as a function of annealing temperature, which is known to convert sp$^{3}$ hybrids to six-fold aromatic domains.  Figure \ref{fig:SEY-CvsT} shows that a moderate structural quality of the carbon layer is sufficient to achieve a considerable SEY decrease as aromatic clusters of limited size approach the secondary emission properties of graphite. This evidence and the ongoing research on different deposition procedures to grow moderate quality amorphous carbon on an industrial scale  suggest that  amorphous carbon currently represent  a potentially very interesting solution for mitigating ECEs. Moreover, Carbon thin films have recently attracted some interest for the accelerator community, due to its capability to reflect Synchrotron Radiation. This Properties could be used to transport SR induced  heat load from cold surfaces in dipoles to warm part of the machine, where it could be more readily dissipated \cite{Cimino2015}. 

Clearly, such coatings also risk to losing their e$^-$ cloud mitigation properties if contaminants  are physisorbed on their surface, so that extra care has to be taken to keep such low SEY coatings  as clean as possible.

\subsection{Solenoid field}
The use and effectiveness of a solenoid field in reducing ECEs in the LHC are shown in Fig. \ref{fig:Mergedvac}.
Unfortunately, solenoids cannot be adopted in many parts of the ring, not only where space constraints do not allow any coil to be wrapped
around the chamber but also because external solenoid fields are not effective in suppressing the build-up of the electron cloud  in magnetic field regions. This make solenoids an effective countermeasure but not sufficient to guarantee total suppression of ECEs.

\subsection{Electrodes}

\begin{figure}
\centering\includegraphics[width=.95\linewidth]{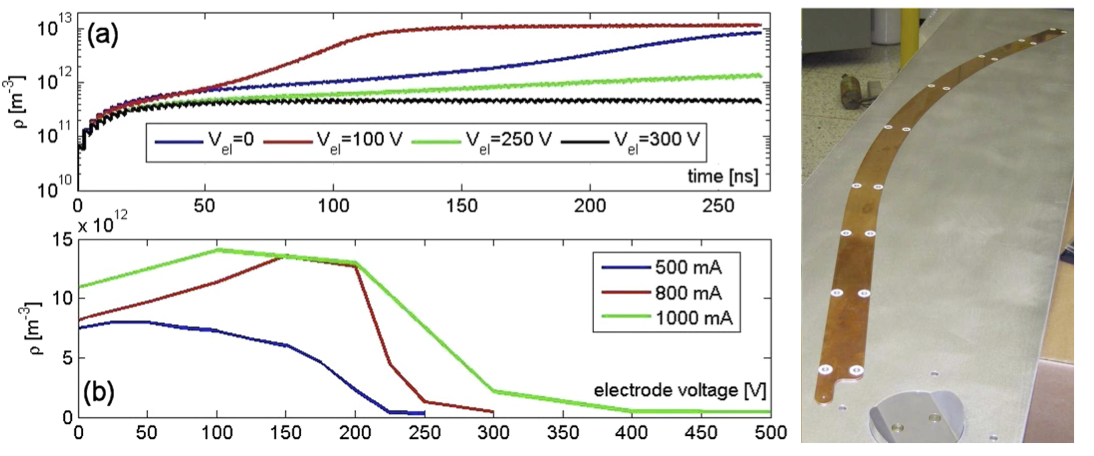}
\caption{(a) Evolution of the averaged cloud density for different values of electrode voltage. (b) e$^-$ cloud density at the end of the bunch train. Right panel: Electrode inserted in the dipole chamber of DA$\Phi$NE\cite{alesini} (courtesy of D. Alesini).}
\label{fig:DAFNE}
\end{figure}

A complementary and very efficient additional active mitigation solution is the use, where possible, of electrodes. This is because the electrons of the cloud can easily drift from one side of the chamber's surface to the other side under the force of a clearing field. In the DA$\Phi$NE collider, an electrode has been proposed and installed in all positron ring arcs\cite{alesini}. In Fig. \ref{fig:DAFNE} (right panel) a picture of such an electrode mounted in the machine dipoles is shown. Figure \ref{fig:DAFNE}(a) shows the e$^-$ cloud density evolution for different values of the electrode voltage for 800\,mA in 100 consecutive bunches. A non-monotonic dependence of the saturation density on the electrode voltage is clearly observed. In particular, the maximum value is reached at around 150\,V. For higher voltages, the density sharply decreases; at 300\,V, it is reduced by about two orders of magnitude. The same behaviour is observed for different values of the beam current, as shown in Fig. \ref{fig:DAFNE}(b).
Of course, electrodes may induce impedance problems, which should be carefully considered before adopting electrodes to mitigate ECEs. In most accelerators, it could not be easy to find space to accommodate them without significant extra cost. A non-trivial technical problem is to guarantee electrical insulation during operation and avoid accumulation of dust.

\subsection{Geometrical modification}
\begin{figure}
\centering\includegraphics[width=.95\linewidth]{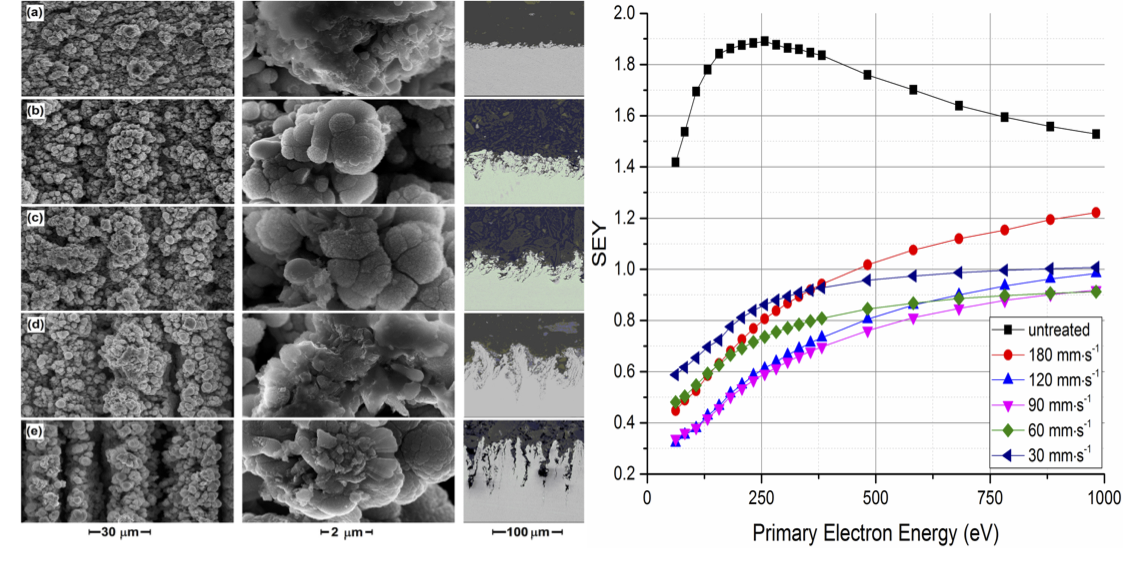}
\caption{Left: Low (left) and high (middle) resolution planar and cross-section (right) scanning electron micrographs of 1 mm thick copper samples treated with lasers at different scan speeds: (a) 180 mm/s, (b) 120 mm/s, (c)  90 mm/s, (d) 60 mm/s, (e) 30 mm/s. Right: SEY of laser-treated samples (a)--(e) with varying scan speed as a function of primary-electron energy and non-treated sample\cite{Reza2017}.}
\label{fig:LES}
\end{figure}

The idea behind any geometrical reduction of SEY is simply based on the experimental fact that electrons contributing to the SEY have an energy distribution peaked at very low kinetic energy, an angular distribution with its maximum at normal emission and a significant decrease at grazing angles.  So, if only electrons emitted from the surface at grazing incidence `see' the beam, and the ones emitted at normal incidence are shadowed by the surface itself, than a net  reduction in the SEY is expected. Macroscopic grooved structures\cite{Kra2004,Pivi2004}, as well as microscopically modified surfaces  \cite{Montero2013,Reza2017}, have been proposed and studied and  their SEYs have been measured, confirming predictions.
Figure \ref{fig:LES} shows the most recent results obtained by laser irradiation
or ablation with different laser working parameters of some Cu surfaces. In the shown case, the use of different exposition power and time varies the resulting surface modifications, as seen by the scanning electron micrograph in Fig.\ref{fig:LES}, giving different SEY curves, all with significantly lower overall values than the untreated Cu.
Such geometrical modifications are indeed shown to be very efficient and promising methods of mitigating ECEs. Great effort is currently made to assess  their effect on beam impedance, to quantify and reduce possible particle generation, and to see the effect on overall vacuum performances of  the significantly increases in the effective surface of such modified structures, as compared with flat surfaces.

\subsection{Electron and photon scrubbing}

\begin{figure}
\centering\includegraphics[width=.95\linewidth]{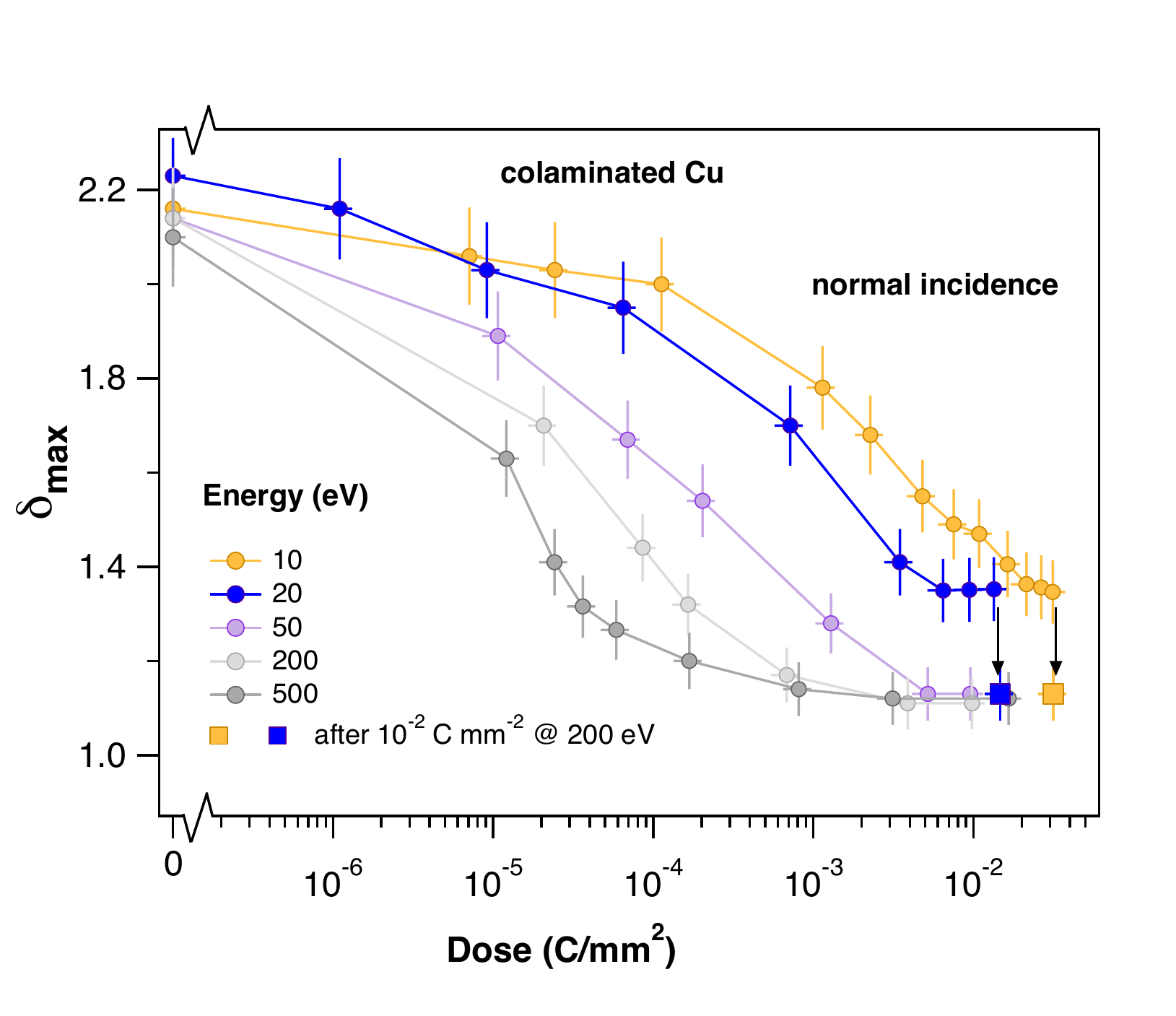}
\caption{$\delta_{\max}$ as a function of dose for different impinging electron energies at normal incidence on colaminated Cu of the LHC beam screen (Higher energy corresponds to lower  $\delta_{\max}$). The squares represent the $\delta_{\max}$ values measured after an additional electron dose of 1.0$ \times $10$^{-2}$\,C/mm$ ^{2} $ at 200\,eV\cite{Cimino2012}.}
\label{fig:scrubbing}
\end{figure}

As already said, most, if not all, materials used in accelerator technology have a significantly high SEY, much higher than that needed to avoid electron-cloud-related instabilities.  Pioneering work done at CERN in this context \cite{ECLOUDProc1, ECLOUDProc2, ECLOUDProc3, ECLOUDProc4, ECLOUDProc5,baglin1998,Cimino2012,Cimino2014} demonstrates that, when a surface is exposed to an electron beam, its SEY decreases. This is shown in  Fig.  \ref{fig:scrubbing}, which reports the SEY values of a Cu technical surface after subsequent electron dose exposures. Such an observation had and still has a profound impact for cloud-related mitigation studies, since electron bombardment is what is actually happening during e$^-$ cloud formation. It appears that, while the cloud is forming, it is able to cure its detrimental effects: the electron beam hits  the accelerator walls and causes a  reduction of their SEY value. Such a process, called `scrubbing', shows a kind of self-mitigation mechanism, first proposed  and then successfully adopted \cite{ECLOUDProc1, ECLOUDProc2, ECLOUDProc3, ECLOUDProc4, ECLOUDProc5} as the  baseline design to cope with any detrimental effects of  electron-cloud-related instabilities in
the  LHC.

\begin{figure}
\centering\includegraphics[width=.9\linewidth]{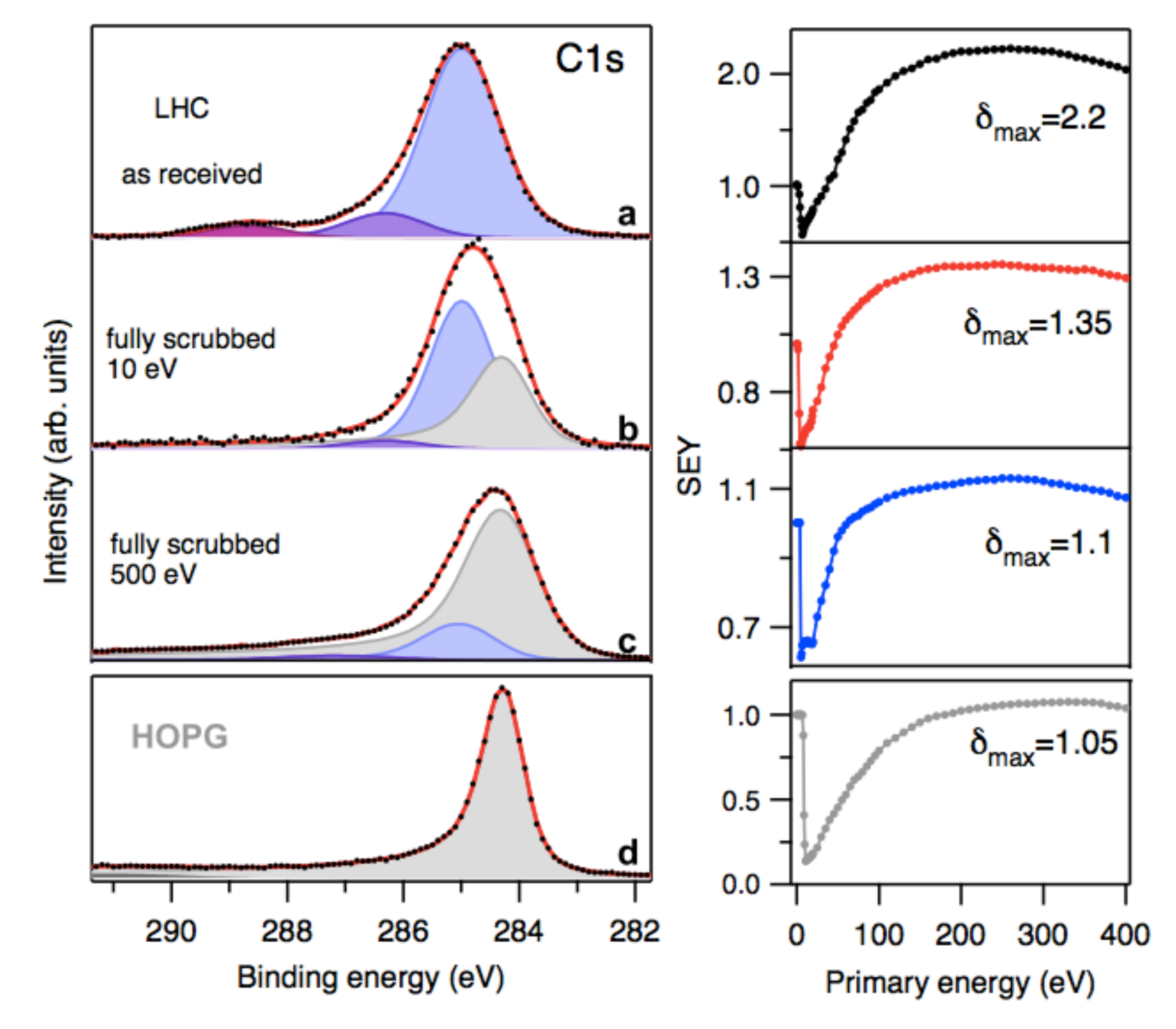}
\caption{Left: C1s X-ray photoelectron spectroscopy spectra (left panel). Right: SEY curves measured on the LHC Cu sample: (a) `as received'; (b) after a dose of  3$\times10^{-2}$\,Cmm$^{-2}$ at 10\,eV;  (c) after a dose of 3$\times10^{-2}$\,Cmm$^{-2}$  at 500\,eV; (d) on a freshly cleaved highly oriented pyrolytic graphite surface \cite{Cimino2012}.}
\label{fig:XPS&SEYCu}
\end{figure}

Figure \ref{fig:scrubbing} also shows that scrubbing depends not only on the electron dose but also on the energy of the impinging electrons; this observation triggered the search to understand the detailed chemical origin at the base of the scrubbing process\cite{Cimino2012, Cimino2014}.
Those studies have demonstrated that the beneficial effect of electron beam scrubbing on these surfaces coincides with the formation of a graphitic surface film and, since the SEY of graphite, and carbon-based materials in general, is less than that of air-exposed metals, the presence of the amorphous carbon thin film reduces the effective SEY of the surface. Graphitic film growth occurs because, in general, technical surfaces are covered by carbon-containing contaminants that, once exposed to the electron flux, tend to decompose and partly rearrange in graphitic assemblies.  The experimental evidence for this phenomenon is shown in Fig. \ref{fig:XPS&SEYCu}, which reports SEY and X-ray photoelectron spectroscopy
measurements of a Cu prototype of the beam screen adopted for LHC \cite{Cimino2012}.

The most striking changes occurring at the surface, as seen by X-ray photoelectron spectroscopy, are exhibited by the C1s core level spectra reported  in Fig. \ref{fig:XPS&SEYCu} (left panel), together with  the relative SEY curves (right panel), for the three representative cases: (a)  the `as-received' surface and the surface fully scrubbed at (b) 500\,eV and (c) 10\,eV. The bottom panels of Fig. \ref{fig:XPS&SEYCu} show, for comparison,   the C1s core level spectrum measured on a freshly cleaved highly oriented pyrolytic graphite, together with its SEY curve.
The observed chemical modifications are directly connected to the SEY reduction reported in the right panel of Fig. \ref{fig:XPS&SEYCu}. While electron irradiation at 500\,eV modifies the chemical state of  almost all the contaminating carbon atoms, producing a  graphitic-like layer coating the copper surface, one notices how the C1s spectra, taken  after a dose of  3$\times10^{-2}$\,Cmm$^{-2}$ at 10\,eV (curve b) in Fig. \ref{fig:XPS&SEYCu} reveal a limited sp$^{3}$--sp$^{2}$ conversion and a consequently limited scrubbing efficiency for the low-energy beam, reaching a stable final SEY value of $\delta_{\max}=1.35$.

Photon irradiation has also been observed to scrub \cite{Cimino1999,Cimino2014}. This effect, rather than that due to the direct interaction between photons and contaminants, is expected to be mediated by the low-energy electrons photoemitted by the solid.
 This simple reasoning supports the notion that photons do scrub,  if less efficiently than energetic electrons, and  suggests that their induced surface modifications are, overall, very similar to those induced by electrons with energies below 20\,eV. Still,  conclusive analysis of the interplay between the SEY and photoelectrons, yielding modification by photons or e$^-$,   is still missing and would be very useful to understand and predict ECE-related behaviours in accelerators or systems where both phenomena can occur.
As shown here, scrubbing is a constantly occurring phenomenon, which is self-limiting, since any SEY reduction causes ECEs to decrease and so the number of electrons hitting the wall and inducing scrubbing. A detailed understanding is certainly required to predict the operational stability and evolution of any surface subject to ECEs. In this context, it is important to say that any exposure to air of the scrubbed surface will require some reconditioning to reproduce the low SEY obtained before venting. This phenomenon, together with the time necessary to obtain  the lowest achievable SEY during operation, seems to confirm the need to find a solution to mitigate ECEs more controllable than
scrubbing.

\section{Conclusions}

The interaction between beams and vacuum system walls  is a complex phenomenon with many detailed and different aspects. Among them, the formation of an electron cloud in the vacuum vessel of an accelerator is shown to be  a major  issue and this is why it is still a fast evolving research field. This study requires synergic competence in different research branches,  from detailed beam dynamics, to experimental studies performed in standard vacuum and accelerator laboratories as well as research using the most sophisticated surface analytical tools and state-of-the-art surface-coating deposition facilities.

I feel that the ultimate solution and a complete understanding of all aspects connected to ECEs are still not available and that more multidisciplinary work and effort are still needed; I  invite interested readers to join this very challenging and rapidly evolving research field.

\section*{Acknowledgements}

I am indebted to the entire community who is working or did work on electron cloud issues in past years.   Special thanks goes to my colleagues and friends Francesco Ruggiero (1957--2007) and Theo Demma (1972--2014), who deeply influenced me, with their scientific activity but also on a personal basis. \\
I acknowledge support  from the European Unions Horizon 2020 Research and Innovation Program under Grant 654305, the EuroCirCol Project, and the INFN Group V MICA project.
%
%

\end{document}